\documentclass[twocolumn,amssymb,nobibnotes,aps,prapplied,superscriptaddress]{revtex4-2}
\usepackage{graphicx}
\usepackage{textcomp}
\usepackage{amsmath}
\usepackage{commath}
\usepackage{color}

\begin{document}
	\title{Topological Flexural Modes in Polarized Bilayer Lattices} 
	\author{Mohammad Charara}
	\affiliation{Department of Civil, Environmental, and Geo- Engineering, University of Minnesota, Minneapolis, MN 55455, USA}
	\author{Kai Sun}
	\author{Xiaoming Mao}
	\affiliation{Department of Physics, University of Michigan, Ann Arbor, Michigan 48109, USA}
	\author{Stefano Gonella}
	\email{sgonella@umn.edu}
	\affiliation{Department of Civil, Environmental, and Geo- Engineering, University of Minnesota, Minneapolis, Minnesota 55455, USA}
	
	\begin{abstract}
			Topological lattices have recently generated a great deal of interest based on the unique mechanical properties rooted in their topological polarization, including the ability to support localized modes at certain floppy edges. The study of these systems has been predominantly restricted to the realm of in-plane mechanics, to which many topological effects are germane. In this study, we stretch this paradigm by exploring the possibility to export certain topological attributes to the flexural wave behavior of thin lattice sheets. To couple the topological modes to the out-of-plane response, we assemble a bilayer lattice by stacking a thick topological kagome layer onto a thin twisted kagome lattice. The band diagram reveals the existence of modes whose out-of-plane character is controlled by the edge modes of the topological layer, a behavior elucidated via simulations and confirmed via laser vibrometer experiments on a bilayer prototype specimen. These results open an alternative direction for topological mechanics whereby flexural waves are controlled by the in-plane topology, leading to potential applications for flexural wave devices with engineered polarized response.
			
		\vspace{0.4cm}
	\end{abstract}
	
	\maketitle

\section{Introduction}
Maxwell lattices are a special class of lattices characterized by having an equal number of degrees of freedom and constraints \cite{maxwell1864calculation}, thus being on the verge of mechanical instability \cite{Lubensky_2015PhonElCritCoordLat, MaoLubApproxNearIsoKagLat, MaoLubMacLatTopoMech_Review, SouslovLub2009IsoPeriodLat}. A prototypical example in two dimensions is the kagome lattice, which has garnered considerable attention for its mechanical properties \cite{fleck_DamTolElBrit,fan_SandPanKagLat, Digby_ImperfSenIsoElaLat, zhang_MechPropNovelPlanLatStruct}, tunable wave-propagation characteristics \cite{Phani_WavePropPerLat, Schaeffer_WavePropReconfigMagElKagLat, Riva_TunableIPTopoEdgeWavKagLat, Chen_ElQuantSpinHallKagLat, gonella2020symmetry}, and for its potential for reconfigurability and property tuning \cite{kapko2009collapse,gao2018two}.
Some of the most interesting phenomena arise at the boundaries of finite lattice domains \cite{SunTwKagLat,MaRhim_LoopStatesEdgeKag,Susstrunk_HelicalEdgeMechTI}. Certain edge properties are conceptually analogous to those observed in electrical or quantum systems, such as topological insulators \cite{QiZhang_TopoInsSupCond, HasanKane_TopoInsula, ma2019valley, rocklin_WeylModesMaxLat}. 

\begin{figure}[t]
	\centering
	\includegraphics[width=0.5\textwidth]{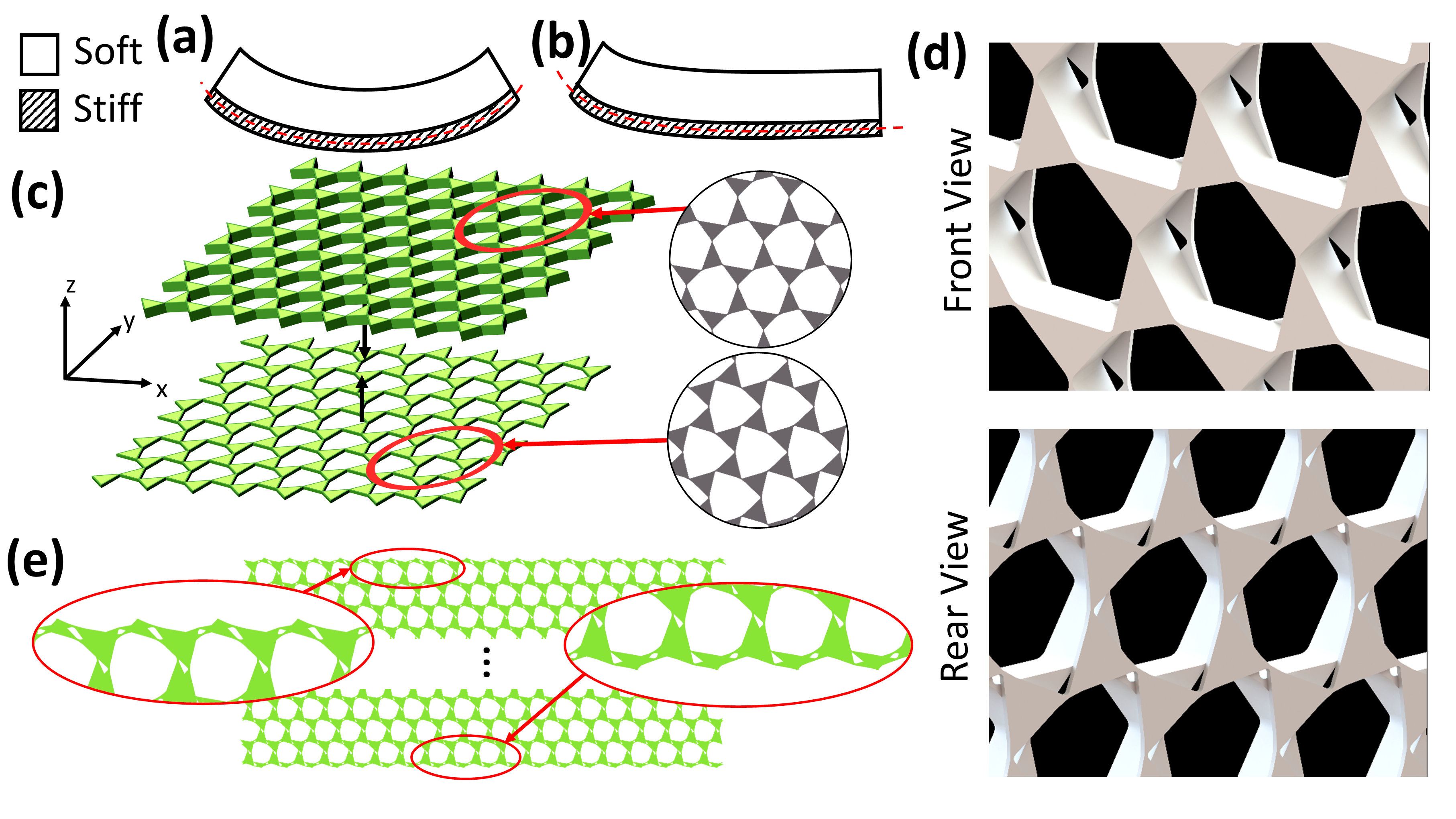}
	\caption{(a) Schematic of coupling-induced bending in a bilayer structure due to strain in the soft layer; (b) localized bending due to excess of softness on one edge of the soft layer; (c) bilayer lattice assembly strategy; (d) close-up details of the front and rear face of the lattice; (e) details of the top and bottom edge.}\label{fig1}
\end{figure}

Recently, a subclass of Maxwell lattices has been shown to exhibit topological behavior \cite{kane2014topological, Kedia_SoftTopoModesRigMechMat, Lubensky_2015PhonElCritCoordLat}, including the ability to localize deformation at a floppy edge in the form of zero-frequency floppy modes, leaving the opposite edge rigid. Although the topological properties manifest at the edges, they are, in fact, intrinsic to the bulk, a property known as bulk-edge correspondence. Further, the topological behavior has been shown to be tunable through a cell reconfiguration, obtained by subjecting the lattice to global zero-energy soft strains, which can cause phase transitions between polarized and nonpolarized states \cite{rocklin2017transformable}. Analogous phenomena have also been realized in structural lattices, an exercise that requires relaxing the perfect hinges to finite-thickness ligaments that can support in-plane bending \cite{ma2018edge, StenLub_TopoPhononSupIsoLat}. As a result, the floppy modes have been shown to rise to finite frequencies, but the signature of polarization and the ability to localize wave modes asymmetrically persist. 
\\
\indent Thus far, the investigation of the topological behavior of two-dimensional (2D) Maxwell lattices has been predominantly restricted to their native in-plane mechanics, without exploring broader implications for their out-of-plane response. In this work, we attempt to close this gap by answering, through a combination of simulations and experiments, the following questions. Are the topological modes exportable to the out-of-plane realm? If so, with what degree of dilution? What is the interplay between flexural bulk and edge modes and how can we distill the signature of the out-of-plane edge modes from complex multimodal wavefields? Note that, while a number of extensions of topological mechanics to three-dimensional (3D) domains have been proposed \cite{Baardink_LocalSoftnessAlong3DTopoMeta, Stenull_TopoPhonWeylLines3D, Bilal_IntrinPolarElMetaMat}, our focus here is on the out-of-plane response of systems whose periodicity remains strictly two-dimensional.
\\
\indent To promote coupling between in-plane and out-of-plane response in a thin structure, we need to establish some degree of nonuniformity through its thickness, which can be obtained by modulating either the material properties or the geometric characteristics, and/or by imposing an eccentric loading stimulus. In a lattice, a natural avenue to modulate the equivalent stiffness is by varying the unit-cell geometry. This modulation can be either continuous through the thickness (as in a functionally graded structure), or piecewise uniform (as in a laminate). Consider, for instance, a bilayer beam obtained by stacking a soft layer on a thinner and stiffer layer. Intuitive structural mechanics considerations suggest that an in-plane state of strain established in the soft layer would result in coupling to an out-of-plane (flexural) deformation experienced by the entire bilayer [Fig.\ \ref{fig1}(a)]. The bending kinematics would necessarily emerge from the different ability of the two layers to accommodate in-plane deformation and from the compatibility requirements at their interface. This notion is commonly exploited in the design of bimorph elements for soft robotics \cite{Soft_Bimorphs_2020} and wave control \cite{celli2017wave}, and to excite flexural waves using piezoelectric patches \cite{Giurgiutiu_LambWavExPZT}. 

For a conventional bilayer, in which the mismatch between the layers is established uniformly over the entire domain, the coupling is also experienced globally by the structure. Suppose instead that the state of strain induced in a layer is localized at one edge, as in a floppy mode of a polarized lattice. We can then expect the resulting coupling to display an analogous degree of localization at the same edge [Fig.\ \ref{fig1}(b)], transferring the asymmetry to the flexural response. The remainder of this paper is devoted to demonstrate this conjecture and quantify the signature of this flexural polarization.
 
\section{A bilayer lattice architecture that promotes coupling}

To explore this idea, we propose a bilayer consisting of a polarized topological kagome lattice stacked on a nonpolarized twisted kagome lattice [Fig.\ \ref{fig1}(c)]. The two layers have complementary roles. The topological layer provides the in-plane polarization required to trigger the desired edge behavior. The twisted kagome merely provides some impedance against the edge deformation mechanisms of the topological layer, thus promoting coupling. 
Accordingly, we make the twisted kagome layer as thin as practically realizable to maximize its flexural compliance while maintaining its in-plane stiffness. This also ensures that the in-plane mechanics of the bilayer are predominantly controlled by the topological layer. In our configuration, henceforth referred to as ML Topo90, 90\% of the total thickness is occupied by the topological layer and 10\% by the twisted kagome layer. 
The triangles in the twisted kagome feature $2.35$-$\textrm{cm}$-sides, $0.12$-$\textrm{cm}$-thick hinges, and a twist angle of $16.4^{\textrm{o}}$. The topological layer has two scalene triangles with side lengths $2.82$, $2.45$, and $2.18\, \textrm{cm}$, and $0.12$-$\textrm{cm}$-thick hinges. The total thickness of ML Topo90 is $1.65\, \textrm{cm}$. Isometric close-up renderings of the front and rear faces are shown in Fig \ref{fig1}(d). It is also helpful to define a reference lattice, which we refer to as Topo100, consisting solely of the topological layer of ML Topo90 taken in isolation. Throughout our analysis, we assume all lattices made from acrylonitrile butadine styrene (ABS) (E = $2.14\, \textrm{GPa}$, $\nu$ = $0.35$), although the results are scalable to other materials.

Note that we deliberately amend the lattice edges by trimming the protruding portion of the edge triangles [Fig.\ \ref{fig1}(e)]. The purpose of this correction is to filter out, or minimize, any edge effects that could result from the direct activation of flexural motion of the edge quasidangling protruding elements. The localization resulting from such mechanisms, if established, would have a trivial nature in that it could not be linked to any intrinsic topological polarization of the bulk and would depend on the specific geometric features of the edges. While trivial edge effects are modest in in-plane problems, their flexural counterparts can display large amplitudes, with the possibility to pollute, if not overshadow, the actual topological effects that we seek to observe. This correction virtually eradicates this possibility, thus distilling the contribution to the edge localization and lattice asymmetry that is germane to the topological polarization. Further details and visualization of the trivial effects observed in an untrimmed version of this lattice are discussed in Appendix A.

\begin{figure}[t]
	\centering
	\includegraphics[width=0.47\textwidth]{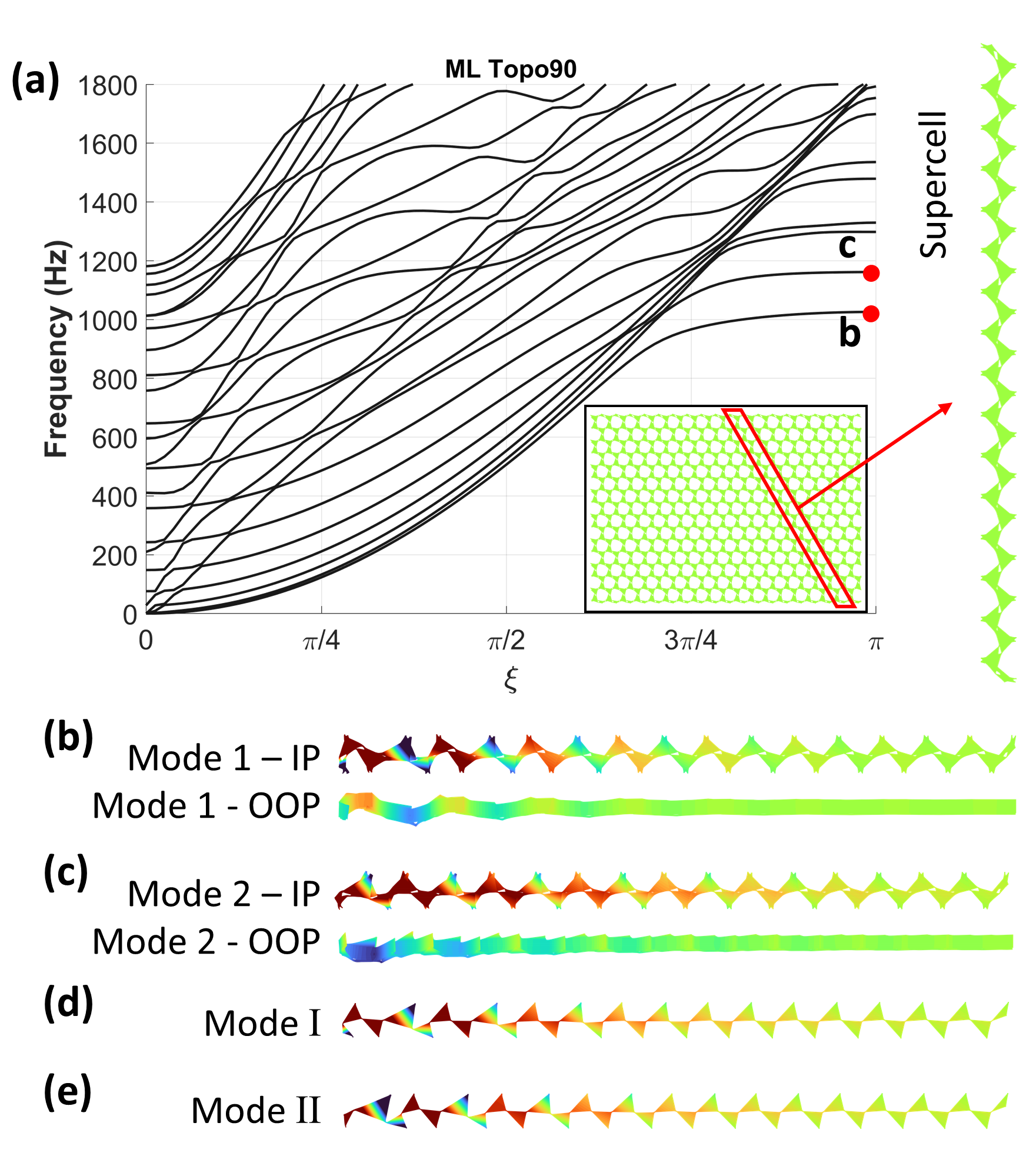}
	\caption{Band diagram of 16-cell supercell for ML Topo90 with mode shapes for the first two modes at $\pi$ (b,c); edge mode shapes at $\pi$ for Topo100 (d,e) shown for reference.}\label{fig2}
\end{figure}

Fig.\ \ref{fig2}(a) shows the band diagram for an ML Topo90 16-cell supercell, modeled using 3D elasticity. The coexistence of in-plane and out-of-plane modes results in a richer and highly hybridized band spectrum compared to the prototypical case of a topological lattice deforming only in-plane, as studied in Ref.\ \cite{ma2018edge}. This additional modal complexity makes the identification of topologically polarized modes more challenging. In the 2D model of a single layer, the topological edge modes live in a frequency interval with low modal density, in which the only other available modes are the in-plane acoustic modes, which can be easily discriminated from the topological ones for their long-wavelength content. In a 3D model of a bilayer, the edge modes co-exist with a plethora of flexural modes with comparable wavelengths. Therefore, their identification cannot rely solely on the spectral characteristics of the branches, but must involve a morphological inspection of the associated mode shapes. 

\begin{figure}[t]
	\centering
	\includegraphics[width=0.5\textwidth]{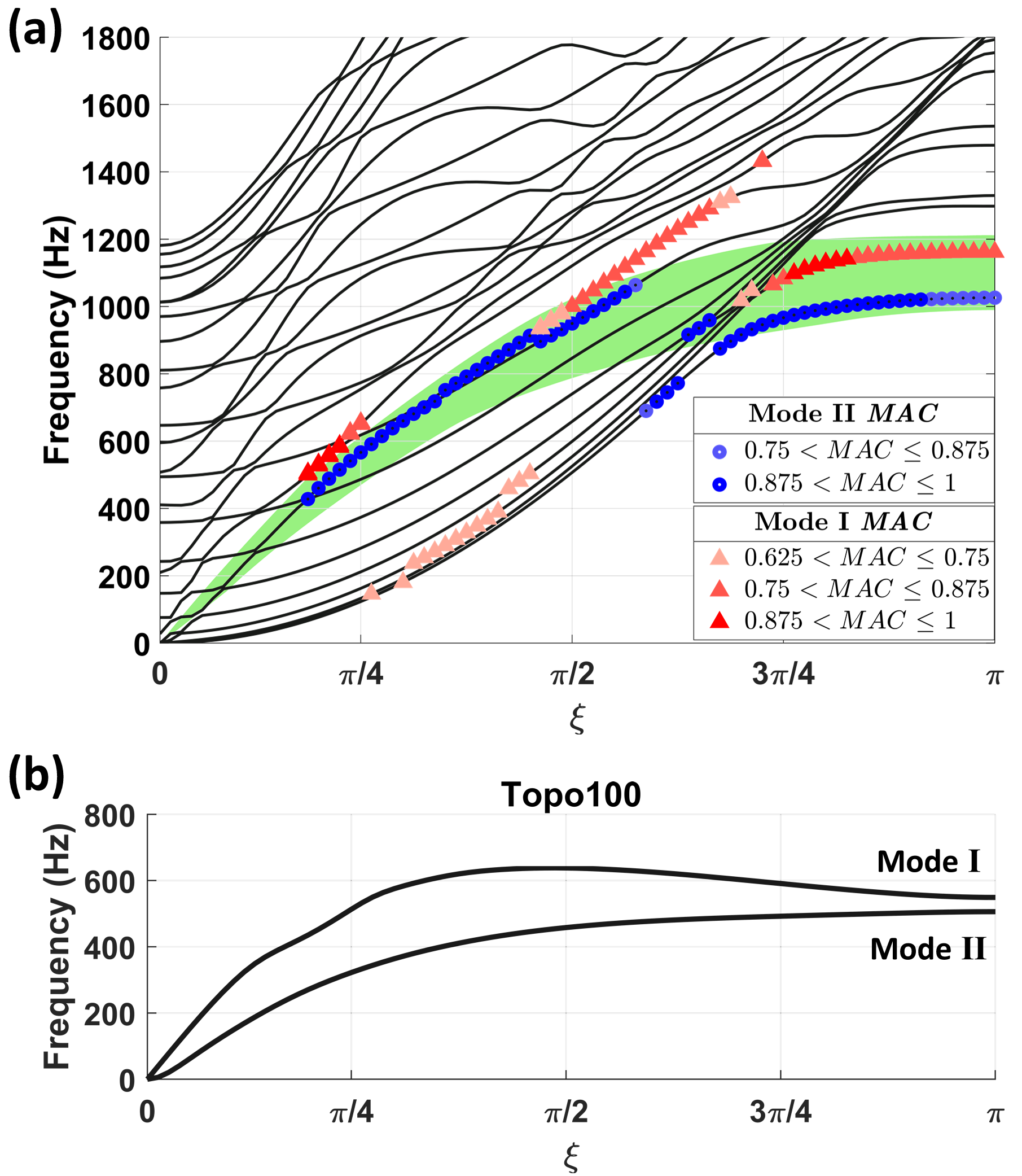}
	\caption{(a) MAC analysis between in-plane deformation of ML Topo90 modes and Topo100 modes (shown in b); the analysis highlights the spectral points with the highest modal correlation, which define a qualitative spectral region of influence in which the bilayer response is dictated by the in-plane behavior of the topological layer.}\label{fig3}
\end{figure}

Here we focus of our attention on modes 1 and 2 of ML Topo90 at $\xi = \pi$, 
whose mode shapes are shown in Fig \ref{fig2}(b) and \ref{fig2}(c), respectively, with in-plane (IP) deformation shown in the top-down view (color proportional to in-plane displacement along \emph{y}) and out-of-plane (OOP) deformation shown in the side view. 
For comparison, the first two branches of the band diagram for Topo100, calculated using 2D elasticity, are reported in Fig.\ \ref{fig3}(b), and their mode shapes at $\xi=\pi$ are plotted in Fig.\ \ref{fig2}(d) and \ref{fig2}(e). We observe that both in-plane and out-of-plane components of the ML Topo90 mode shapes feature a high decay rate at the top (floppy) edge, a clear signature of polarization. Interestingly, the mode shapes of Topo100 display a similar deformation pattern and decay rate, supporting the hypothesis that the flexural response of ML Topo90 imports its polarization attributes from the topological layer through coupling.

To quantify the link between the morphological characteristics of the ML Topo90 modes and their Topo100 in-plane counterparts, we perform a modal assurance criterion (MAC) analysis, where the quantity $MAC=\boldsymbol{|\phi_1 \cdot \phi_2|/(\phi_1 \cdot \phi_1)(\phi_2 \cdot \phi_2)}$ estimates the degree of compatibility between two eigenvectors, $\phi_1$ and $\phi_2$. 
The MAC values vary between 0 and 1, with 1 indicating perfect compatibility and 0 orthogonality. We sweep the band diagram to find the modes of ML Topo90 that display the highest MAC correlation, at each value of $\xi$, to the topological modes of Topo100. 
Figure \ref{fig3} highlights the spectral points on the branches of ML Topo90 whose eigenmodes are most correlated with those of either edge mode of Topo100 (blue and red markers referring to mode I and II, respectively), with color intensity proportional to the strength of correlation. It is evident that, near $\xi=\pi$ , where the first two modes of ML Topo90 are spectrally isolated, the correlation with Topo100 is high, consistent with our visual inspection of the mode shapes in Figs.\ \ref{fig2}(b)-\ref{fig2}(e). As we move away from $\pi$, the identified spectral points are more scattered across the available branches, as expected given the competition of several hybridizing modes at longer wavelengths. This said, the highest compatibility points remain organized along spectral paths that are qualitatively reminiscent of the Topo100 edge-mode branches [Fig.\ \ref{fig3}(b)], and identify a spectral ``region of influence" [whose qualitative support is highlighted in green in Fig.\ \ref{fig3}(a)] in which the response of ML Topo90 is heavily controlled by the topological layer. These observations corroborate the notion that the out-of-plane response of the ML Topo90 modes is dictated by the in-plane topological behavior of the topological kagome layer through coupling. Moving closer to the continuum limit increases the chance of ``false positives" due to the higher morphological similarity between modes at long wavelengths. For these reasons, we restrict our considerations to $\xi > \pi/2$.

\begin{figure}[t]
	\centering
	\includegraphics[width=0.49\textwidth]{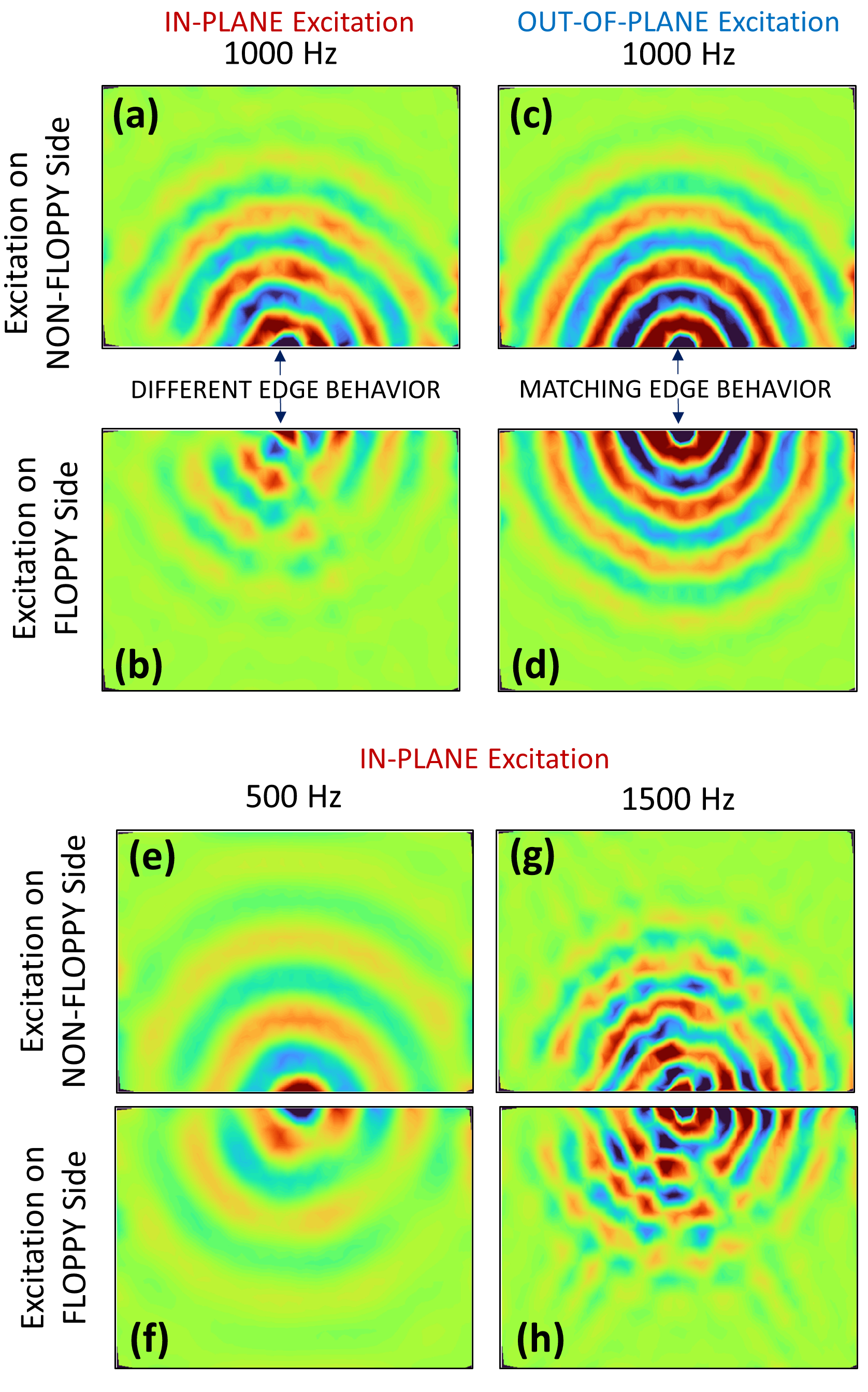}
	\caption{Full-scale simulations. Snapshots of out-of-plane wavefields for in-plane excitation at 500, 1000, and 1500 Hz from the nonfloppy (a,e,g) and floppy (b,f,h) edges; out-of-plane wavefields for out-of-plane excitation at 1000 Hz from the nonfloppy (c) and floppy (d) edges.}\label{fig4}
\end{figure}

\begin{figure}[t]
	\centering
	\includegraphics[width=0.47\textwidth]{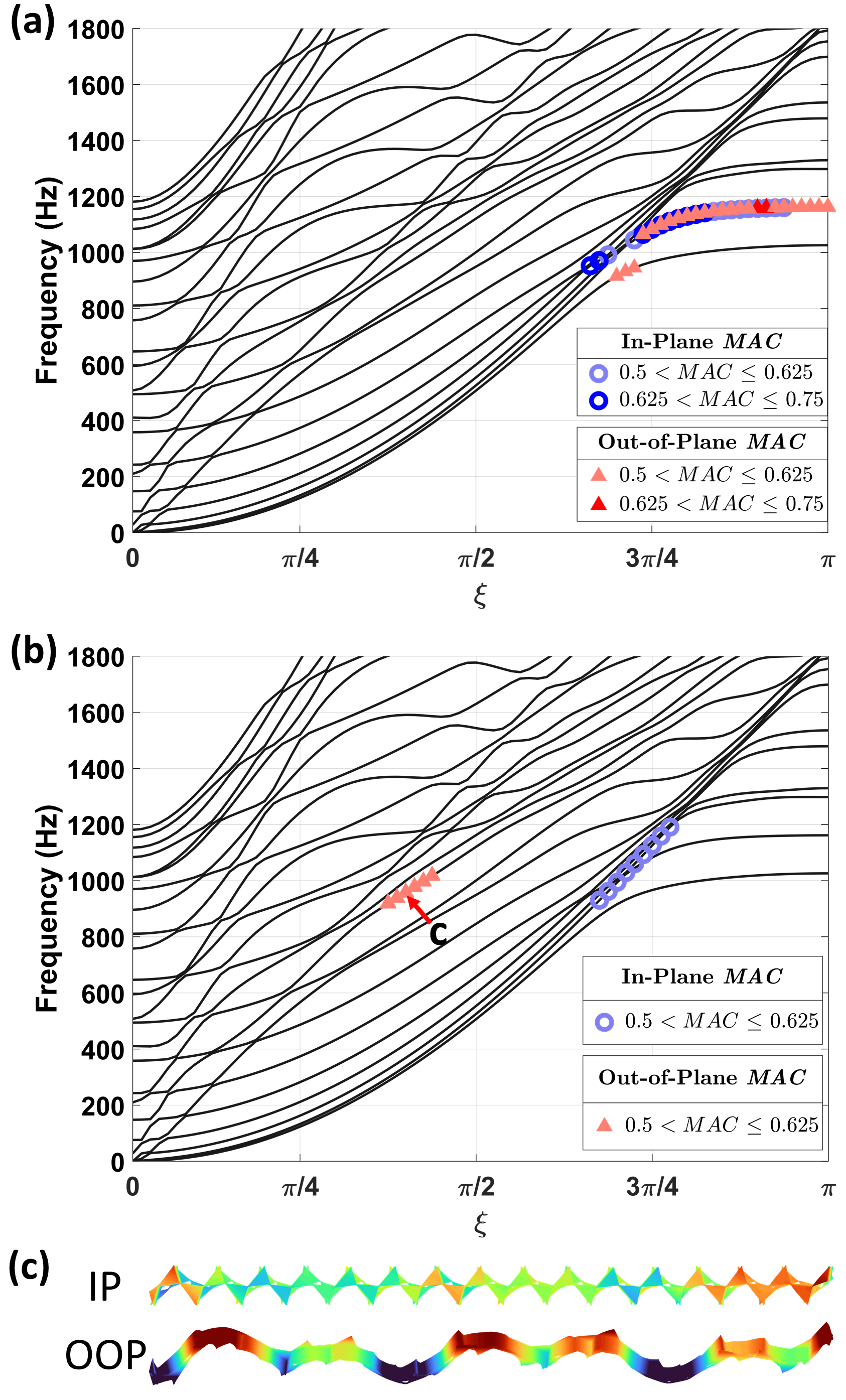}
	\caption{MAC analysis between the response of a strip extracted from the ML Topo90 lattice wavefield excited at $1000\, \textrm{Hz}$ from the floppy (a) and nonfloppy (b) edge and the ML Topo90 supercell modes available in the same spectral interval; (c) example of bulklike modeshapes activated by nonfloppy edge excitation.}\label{fig5}
\end{figure}

\section{Full Scale simulations and the emergence of asymmetric flexural wavefields}
To illustrate the manifestation of these phenomena at the lattice scale we simulate wave propagation in an 8$\times$19 ML Topo90 lattice excited with a five-cycle tone burst. 
In order to maximize the achievable coupling and promote the in-plane localization that we intend to transfer to the flexural modes, we prescribe an in-plane excitation force directly to the plane corresponding to the outer face of the topological layer. Figure \ref{fig4} shows snapshots of a propagating flexural wavefield for excitation prescribed at the top (floppy) and bottom (nonfloppy) edges, respectively, with carrier frequencies of $500\, \textrm{Hz}$ [Figs.\ \ref{fig4}(e) and \ref{fig4}(f)], $1000\, \textrm{Hz}$ [Figs.\ \ref{fig4}(a) and \ref{fig4}(b)], and $1500\, \textrm{Hz}$ [Figs.\ \ref{fig4}(g) and \ref{fig4}(h)]. Excitation at $1000\, \textrm{Hz}$, falling in the interval where the modes display the strongest polarization, reveals strong asymmetry between waves fired from the floppy and nonfloppy edges. Specifically, from the nonfloppy side,  waves travel into the bulk isotropically with the typical circular crest structure of flexural waves, while, from the floppy side, the penetration into the bulk is more impeded and we observe a certain degree of localization along the edge.

\begin{figure}[t]
	\centering
	\includegraphics[width=0.47\textwidth]{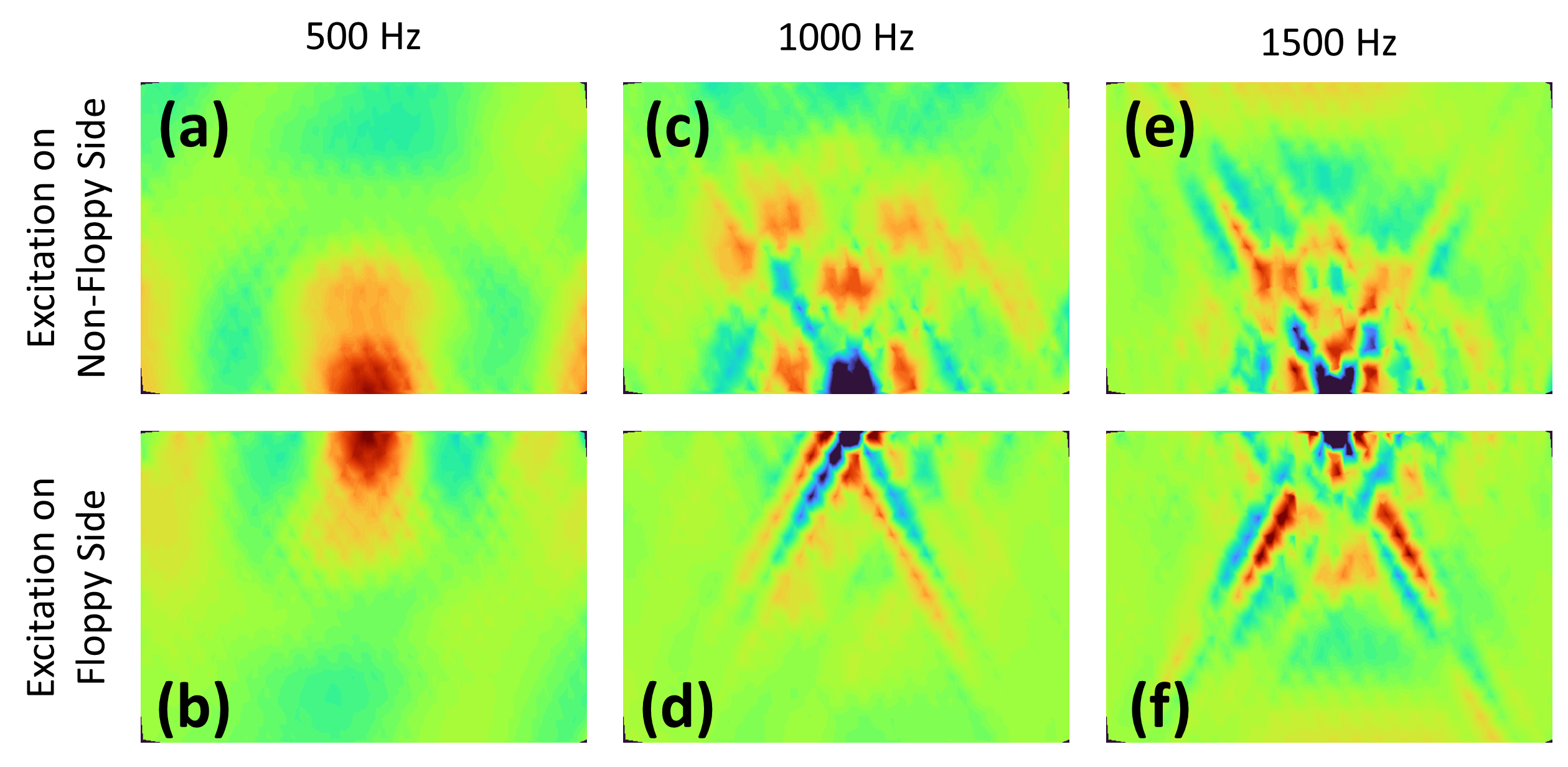}
	\caption{Simulated in-plane displacement wavefields for in-plane excitation from the nonfloppy and floppy side at frequencies of 500 (a),(b), 1000 (c),(d), and 1500 Hz (e),(f), respectively.}\label{fig6}
\end{figure}

It is worthwhile to note that, in this problem, we observe some wave leakage into the bulk even from the floppy edge, resulting in a weaker asymmetry between the edges than the one recorded for in-plane waves in Ref.\ \cite{ma2018edge}. 
This is due to the fact that, in the frequency interval of our burst, the band spectrum features several flexural bulk modes than can be activated in conjunction with the edge modes. This is not the case in the in-plane problem, where the only competition to the edge modes comes from the acoustic \emph{S} and \emph{P} modes. 
In contrast with the $1000$-$\textrm{Hz}$ case, the wavefields for excitation at $500$ and $1500\, \textrm{Hz}$ are significantly more symmetric, 
suggesting that the asymmetric wave transport is frequency selective and is mostly absorbed below and above the frequency interval of the polarized modes. 

The nontriviality of this result can be appreciated by considering the complementary case in which the bilayer is excited with an out-of-plane force. The scope of this test is to confirm that the observed asymmetry is indeed due to topological polarization, and, as such, intrinsic to the bulk and topologically protected, and not a mere consequence of some trivial asymmetry between the local geometric features of the edges. Wavefields for an out-of-plane excitation at $1000 \textrm{Hz}$  from the floppy and nonfloppy edges are shown in Fig.\ \ref{fig4}(c) and \ref{fig4}(d), respectively. The response features circular-crested waves propagating into the bulk, with no edge localization and nearly matching behavior between the edges. If the lattice displayed any trivial edge behavior associated with the protruding edge elements, such effects would be magnified by an out-of-plane force that would directly activate the flexural motion of these elements. 
This is clearly not the case here, suggesting that the asymmetry observed at $1000\, \textrm{Hz}$ [Figs. \ref{fig4}(a) and \ref{fig4}(b)] must originate from the polarization of the in-plane modes.

To pinpoint which modes are responsible for the established asymmetry, we can again resort to MAC analysis.  To this end, from Fig.\ \ref{fig4}(a) and \ref{fig4}(b) we choose a strip of 16 cells spanning the entire lattice 
and we compare its in-plane and out-of-plane displacement fields against the eigenvectors of the ML Topo90 supercell. In Figs.\ \ref{fig5}(a) and \ref{fig5}(b), the markers show the ML Topo90 modes that display the highest modal correlation with the strip deformation at each value of $\xi$ in the neighborhood of $1000\, \textrm{Hz}$, for excitation from the floppy [Fig.\ \ref{fig5}(a)] and nonfloppy [Fig.\ \ref{fig5}(b)] edge. For excitation at the floppy edge, the procedure highlights the two coupled topological modes near $\xi = \pi$. 
In contrast, excitation at the nonfloppy end appears to activate 
bulk flexural modes that lack topological character [Fig.\ \ref{fig5}(c)]. 

Additional insight into the wave-transport characteristics of the bilayer lattice can be gained from the inspection of the in-plane displacement wavefields extracted from the same set of simulations. Fig.\ \ref{fig6} shows displacement wavefields in the \emph{y} direction for excitation at the nonfloppy and floppy edges at carrier frequencies of $500\, \textrm{Hz}$ [Figs.\ \ref{fig6}(a) and (b)], $1000\, \textrm{Hz}$ [Figs.\ \ref{fig6}(c) and (d)], and $1500\, \textrm{Hz}$ [Figs.\ \ref{fig6}(e) and (f)]. At $1000\, \textrm{Hz}$, where the polarized in-plane supercell modes fold, the wavefield displays a marked asymmetry between the edges akin that of Ref.\ \cite{ma2018edge}. Specifically, excitation at the nonfloppy edge results in nearly perfect isotropic wave propagation deep into the bulk, while excitation at the floppy edge features directionality and substantial localization at the edge. Note that also the excitation from the floppy end is not completely immune from a certain degree of leakage into the bulk, likely due to the influence of the twisted kagome layer, which is nonpolarized; however this effect is almost insignificant compared to the bulk wavefield from the nonfloppy edge. 

\section{Experiments on Structural Lattices}

\begin{figure}[t]
	\centering
	\includegraphics[width=0.47\textwidth]{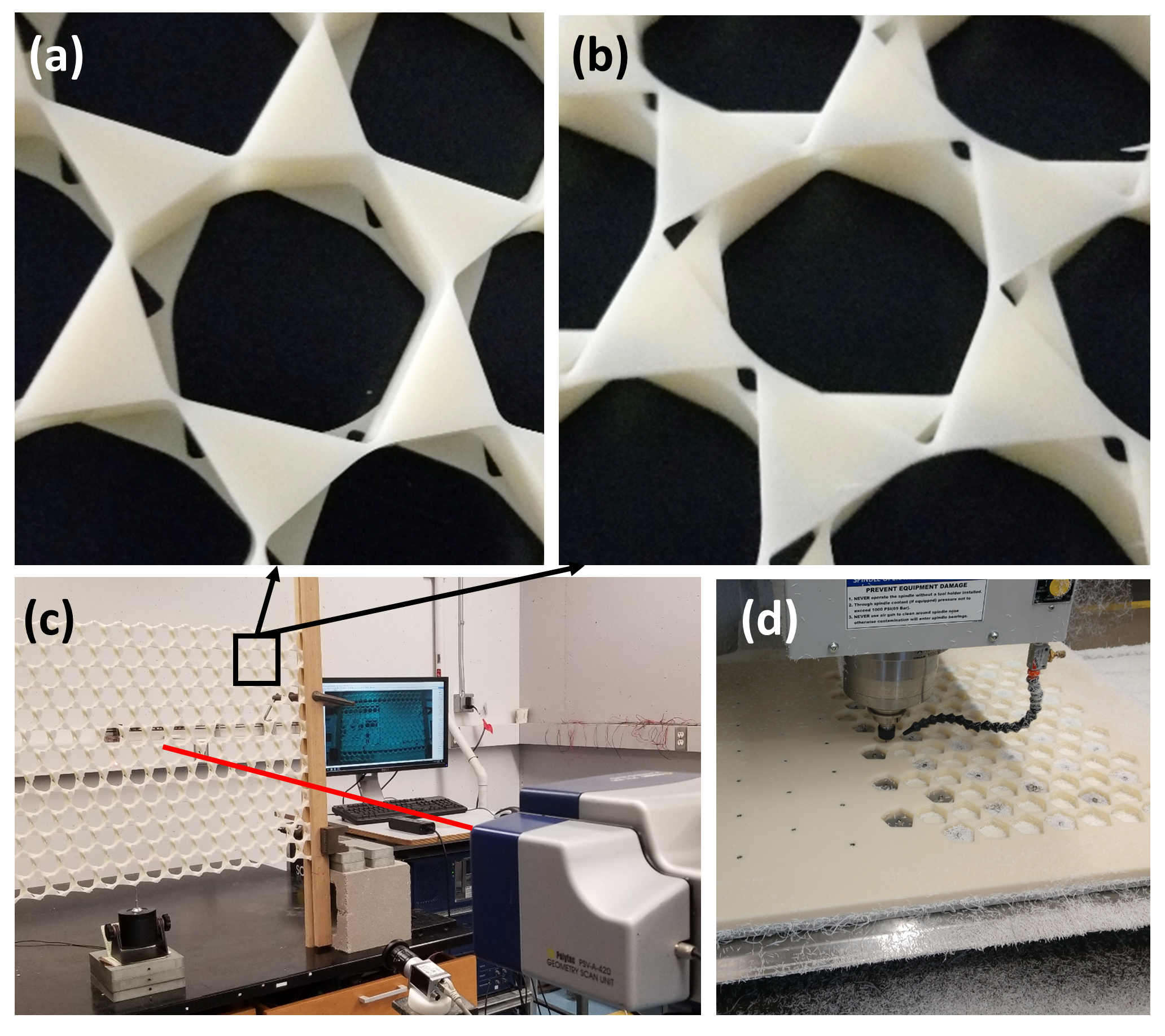}
	\caption{(a) Front and (b) rear detail of the lattice revealing high-precision realization of the hinges in both layers obtained via double-side CNC machining; (b) experimental setup and (c) manufacturing of the bilayer lattice via double-sided CNC machining (cutting of the topological layer shown) .}\label{fig7}
\end{figure}

To verify these phenomena experimentally, we perform laser-assisted (Polytec PSV 400 3D Laser Doppler Vibrometer) experiments on a 7$\times$17-cell ML Topo90 prototype manufactured via double-side computer numerical control (CNC) machining of an ABS sheet; an image of the manufacturing process is shown in Fig.\ \ref{fig7}(d). Due to the complexity of this structure, the machining is done in two steps. First, a plastic plate is fastened to a previously machined manufacturing plate and the twisted kagome layer is milled to 10\% of the total thickness. The structure is then flipped over to the opposing face, reattached to the manufacturing plate, and the topological kagome layer is milled. The result is a bilayer lattice made from one single piece of ABS plastic. The specimen has the notable advantage of featuring identical material properties everywhere. Also, it lacks the interface nonidealities that would inevitably affect any configuration obtained by bonding separate layers, which would derive from regions of debonding or from the inherent mechanical response of the bonding material.

\begin{figure}[t]
	\centering
	\includegraphics[width=0.47\textwidth]{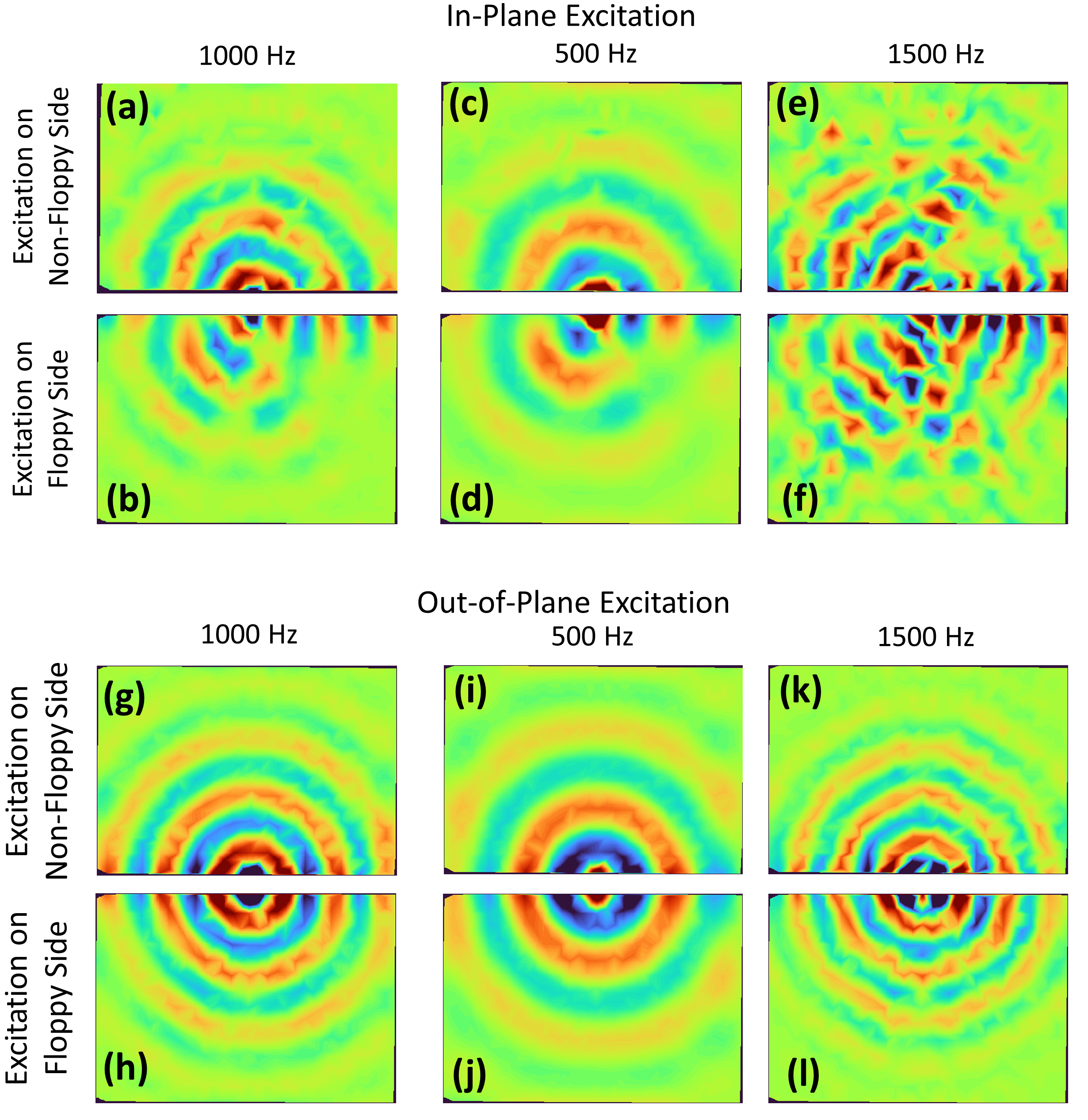}
	\caption{Laser-assisted wave-propagation experiments - out-of-plane wavefield snapshots due to in-plane excitation from the non-floppy and floppy edges at $500$ (a),(b), $1000$ (c),(d), and $1500\, \textrm{Hz}$ (e),(f); out-of-plane wavefield snapshots due to out-of-plane excitation from the  nonfloppy and floppy edges at $500$ (g),(h),$1000$ (i),(j), and $1500\, \textrm{Hz}$ (k),(l).}\label{fig8}
\end{figure}

The excitation is applied by a shaker with the stinger probing the outer face of the topological layer parallel to the lattice plane. The laser scans the lattice surface, measuring the out-of-plane velocity at one point per triangle. The experimental setup is shown in Fig.\ \ref{fig7}(c), with detailed closeups of the specimen in Figs.\ \ref{fig7}(a) and \ref{fig7}(b).  Representative wavefield snapshots for a $1000$-$\textrm{Hz}$ in-plane excitation at the nonfloppy and floppy side are shown in Figs.\ \ref{fig8}(a) and \ref{fig8}(b), respectively, while snapshots for an out-of-plane excitation at the non-floppy and floppy edge are shown in Fig.\ \ref{fig8}(g) and \ref{fig8}(h), respectively. The results match the simulations nicely. The in-plane excitation produces an asymmetric response, with localization at the floppy side and isotropic propagation into the bulk from the nonfloppy side. In contrast, the out-of-plane excitation triggers flexural bulk waves that are virtually insensitive to the polarization. 

For completeness, we also perform experiments 
at $500$ and $1500\, \textrm{Hz}$. Figure \ref{fig8} shows wavefield snapshots for in-plane excitations from the floppy and nonfloppy edges at carrier frequencies of $500\, \textrm{Hz}$ [Figs.\ \ref{fig8}(c) and \ref{fig8}(d)] and $1500\, \textrm{Hz}$ [Figs.\ \ref{fig8}(e) and \ref{fig8}(f)], and for out-of-plane excitation from the floppy and nonfloppy edges at $500\, \textrm{Hz}$ [Figs.\ \ref{fig8}(i) and \ref{fig8}(j)] and $1500\, \textrm{Hz}$ [Figs.\ \ref{fig8}(k) and \ref{fig8}(l)]. The flexural wavefields show that flexural excitation induces only isotropic waves propagating into the bulk regardless of the excited edge, a result confirming the key finding that the in-plane mechanics are a required mediator to trigger the desired topological behavior. However, somewhat surprisingly, we find that the asymmetric behavior experienced by the lattice for in-plane excitation at $1000\, \textrm{Hz}$ persists at $500$ and $1500\, \textrm{Hz}$ Hz, albeit in a much weaker form, especially for the latter. This may be due to a number of imperfections, which are inevitable in an experimental setting, that may cause it to deviate from its numerical counterpart. Specifically, there may exist a slight spectral mismatch between the lattice used in simulations and the actual prototype due to some meshing shortcomings in our computational model, which we could curb, but not fully remove, within the limitations of our finite-element platform. Another factor could be the much lower number of measurement points used in the experiment (two scan points per unit cell) versus the simulation (about 1100 nodes per unit cell). These discrepancies require more investigations and will be further studied in our future work.

 It is worth mentioning that we can consider a variety of alternate avenues to promote coupling between the layers. In Appendix B, for the sake of example, we report the analysis of a bimaterial configuration featuring geometrically identical layers, in which the mismatch is promoted solely by the moduli mismatch between the layers. This solution offers a qualitatively similar response, suggesting that the physics reported in this study may have a fairly universal applicability, beyond the specifics of the selected configuration.

\section{Conclusions}
In conclusion, by designing a bilayer with in-plane to out-of-plane coupling capabilities, we demonstrate the ability to export topological modes from their native in-plane elastodynamics to the flexural realm. These results open the doors to the application of topological wave-manipulation strategies to thin structures traditionally designed to operate in bending. The study reveals the role of the in-plane topologically polarized modes as an essential mediator between a prescribed excitation and the onset of polarization effects in the flexural response.

\begin{acknowledgments}
	
This work is supported by the National Science Foundation (NSF Grant No. EFRI-1741618). We acknowledge the help of Peter Ness at the UMN CSE shop for his expert assistance with the specimen fabrication.
\end{acknowledgments}
	
\appendix

\section{\uppercase{Trivial edge modes in a bilayer topological lattice prior to edge clipping}}

\begin{figure}[t]
	\centering
	\includegraphics[width=0.47\textwidth]{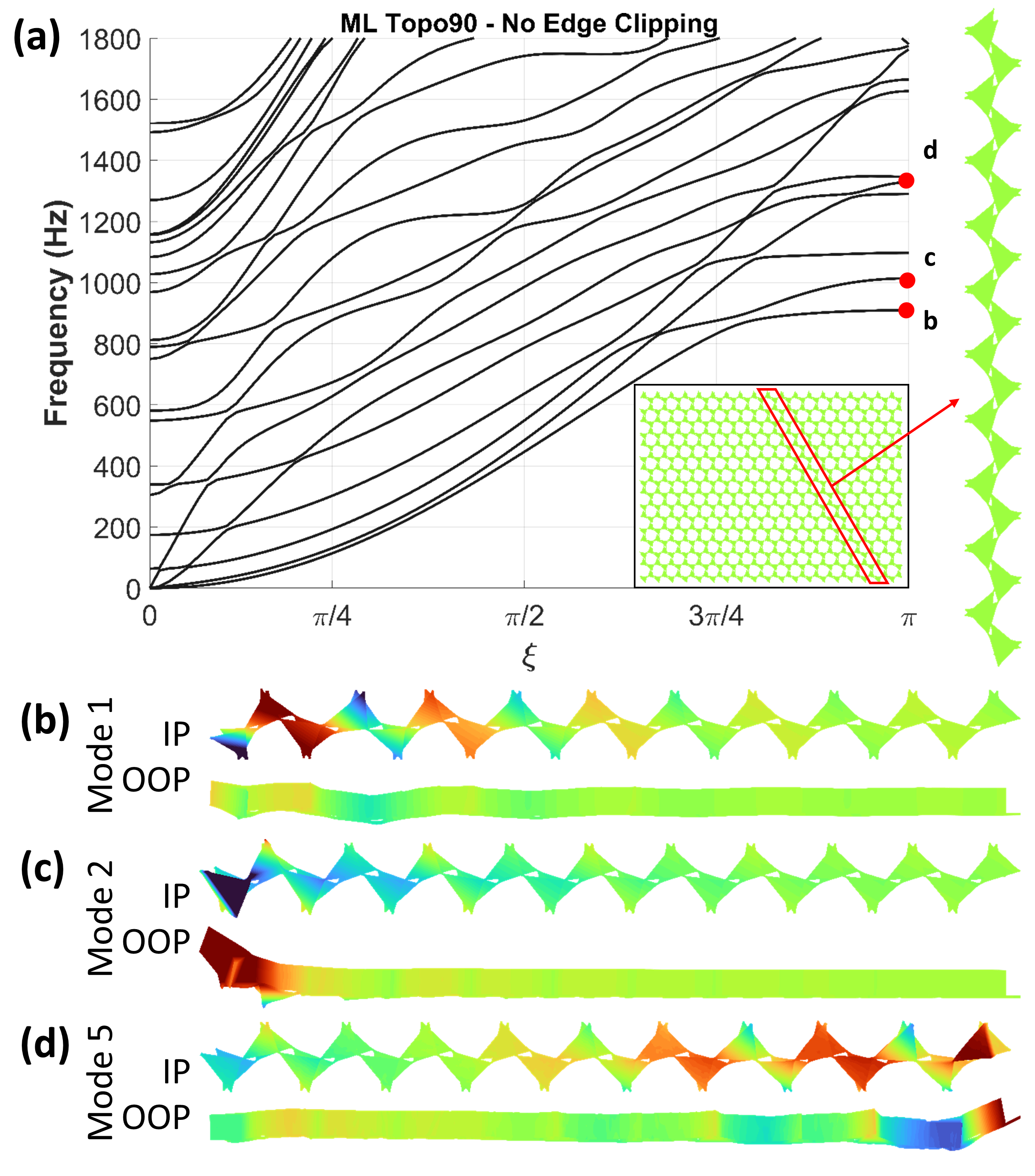}
	\caption{(a) ML Topo90 band diagram prior to edge clipping, with lattice in the bottom right corner inset and supercell highlighted; in-plane and out-of-plane displacement for modes 1 (b), 2 (c), and 5 (d) of ML Topo90 prior to edge clipping.}\label{fig9}
\end{figure}

To demonstrate how the establishment of trivial flexural edge behavior would be detrimental to our ability to extract the signature of topological polarization, in Fig.\ \ref{fig9}(a) we show the band diagram of the lattice prior to edge clipping. In-plane displacement (with color proportional to displacement in the \emph{y} direction) and out-of-plane displacement (with color proportional to out-of-plane displacement) for modes 1, 2, and 5 at $\xi=\pi$ are shown in Figs.\ \ref{fig9}(b)-\ref{fig9}(d), respectively. While the behavior of mode 1 is qualitatively analogous to that discussed for the trimmed edge, the interpretation of mode 2 becomes more problematic. Here we see a very aggressive localization of out-of-plane displacement on the left (floppy) edge. Although a localization at this edge is nominally what we want, it is clear that here the localization is concentrated in the last half cell and can be therefore attributed to the quasidangling nature of the protruding element, rather than to the existence of polarization in the bulk. Perhaps of more relevance, if we inspect mode 5 (a mode not discussed in the paper because of its lack of interesting features in the edge-clipped version of this structure) we see that it displays a similar out-of-plane localization on the nonfloppy edge, acting as the trivial twin of mode 2. In addition, the out-of-plane localization is incommensurate to its in-plane displacement pattern, confirming that this flexural behavior is not triggered by topological in-plane modes.

\section{\uppercase{Bimaterial bilayer lattice}}

\begin{figure}[t]
	\centering
	\includegraphics[width=0.47\textwidth]{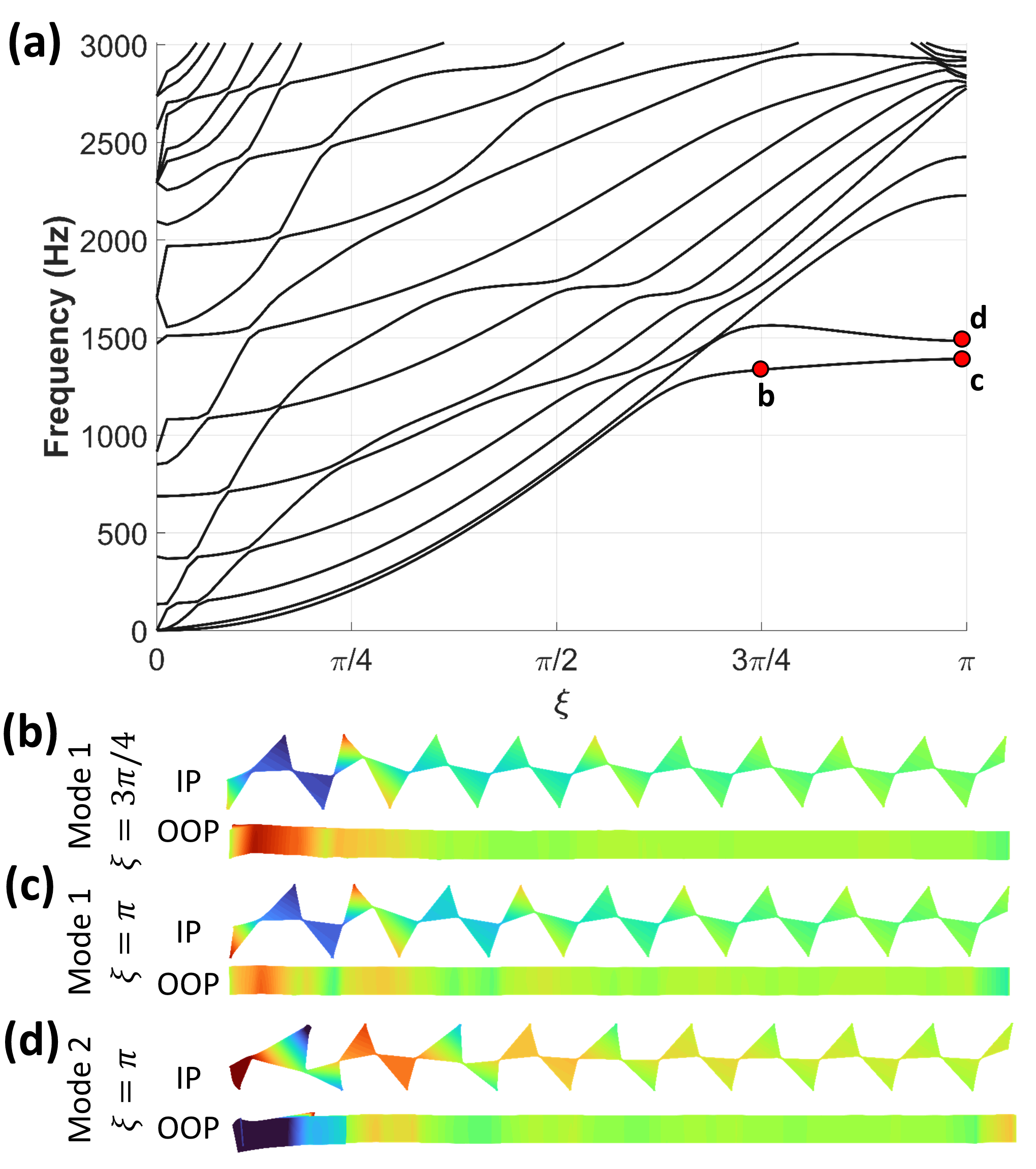}
	\caption{(a) Band diagram: topological kagome - 99\% soft ($E = 2e9$) and 1\% stiff ($E = 2e12$); mode shapes for mode 1 at (b) $\xi=3\pi/4$ and (d) $\xi=\pi$, and (e) mode 2 at $\xi=\pi$.}\label{fig10}
\end{figure}

We consider other strategies to achieve coupling between in-plane and out-of-plane deformation, and transfer floppy-edge behavior to the flexural modes. For the sake of completeness, we report here the results for one of these strategies, which involves a bimaterial platform. Here the layers are geometrically identical and the impedance mismatch between the layers is solely due to the difference between the materials' elastic moduli. The band diagram in Fig. \ref{fig10} refers to a bimaterial topological kagome whose thickness breakdown is as follows: 99\% soft and 1\% stiffer (by 3 orders of magnitude). Note that the specific moduli used here are toy parameters selected for this demonstration with the sole purpose to model a stiffness contrast, and are not meant to be representative of any actual material. Modes 1 and 2 (the two lowest modes) are those associated with the topological edge modes of the topological kagome. Figures \ref{fig10}(b) and \ref{fig10}(c) show the mode shapes for mode 1 at $\xi=3\pi/4$ and $\xi=\pi$, respectively, where the top-down view shows the in-plane deformation (with color corresponding to displacement along y) and the side view captures the out-of-plane deformation (with color corresponding to displacement along z). Both mode shapes show consistency between the decay rates of the out-of-plane and in-plane deformation at the floppy end (left end). Note that the out-of-plane deformation is stronger when we move away from $\xi=\pi$ towards $\xi=3\pi/4$, suggesting that the displacement modes associated with out-of-plane deformation couple best with this edge mode toward the crossing point of the modes, where stronger hybridization occurs. Figure \ref{fig10}(d) shows the mode shape for mode 2 at $\xi=\pi$. Here, the out-of-plane deformation’s decay rate and morphology does not seem commensurate to that of its in-plane counterpart, suggesting that the coupling from in-plane to out-of-plane may include some trivial edge effects not directly associated with the topological character of the in-plane edge mode. In addition, some localized out-of-plane deformation can be seen at the nonfloppy end (right end), further suggesting the presence of a trivial component.

Overall, the bilayer in this paper seems to provide superior coupling performance in terms of preserving features of the topological mode in the flexural response. Nevertheless, we deem this configuration quite effective at localizing flexural waves at the floppy edge, making it a viable alternative to the strategy discussed in the paper and confirming that the main philosophy explored in this paper has universal applicability beyond the specifics of the structure considered in our analysis. 

This said, the configuration studied in this work features one advantage. Working with a single material platform presents major practical advantages in terms of manufacturability and testability. In practice, working with two materials would require either making two lattices and bonding them together \emph{a posteriori}, or 3D-printing the bilayer switching between materials after a few layers of deposition. The first option would require a layer of binding material, which would in turn introduce nonidealities in the response. The second option would be limited by the quality and reliability of the 3D-printed materials currently available, which are generally not satisfactory for dynamical applications. Either way, the structure would exhibit high levels of damping, nonlinearities, material properties uncertainty and heterogeneity, unwanted porosity and scattering, and other nonidealities that would likely result in noisy and elusive measurements. Working with a single material allows us instead to rely solely on more traditional machining (albeit using a nontrivial two-face approach, as discussed above). As a result, we can employ sheets of ABS with very uniform and well-known material properties.

\bibliography{topo3DPaper_bib}

\end{document}